\documentclass{ijcaArticle}
\ijcaVolume{VV}
\ijcaNumber{N}
\ijcaYear{YYYY}
\ijcaMonth{Month}

\ijcaVolume{*}
\ijcaNumber{*}
\ijcaYear{2021}
\ijcaMonth{--------}
\begin{document}

\title{Using Hybrid Scheduling Algorithms For Solving Blockchain Allocation On Cloud} 

\author{
   \large M. A. El-dosuky \\[-3pt]
   \normalsize Computer Science Dep. Faculty of Computer and Info, Mansoura University, Egypt   \\[-3pt]
    \normalsize Computer Science Dep. Arab East Colleges, Riyadh, KSA \\[-3pt]
    \normalsize	meldosuky@gmail.com \\[-3pt]
  \and
   \large Gamal H. Eladl  \\[-3pt]
   \normalsize Information Systems Dep. Faculty of Computer and Info, Mansoura University, Egypt  \\[-3pt]
    \normalsize	gamalhelmy@mans.edu.eg \\[-3pt]
}

\terms{Blockchain, Cloud, Scheduling}
\keywords{Blockchain-as-a-service, Scheduling, shortest-job-first, priority}

\maketitle

\begin{abstract}
Companies are rushing to deliver their services and solutions through the cloud. The scheduling process is very critical in reducing delays.
Scheduling also has a role in accessing resources without excessive waiting time.
All this in context of modern advances in infrastructure and the emergence of Blockchain-as-a-service.
What if integration is done between a hybrid scheduling algorithm and blockchain technology via the cloud. This integration aims to enhance and provide the service uninterruptedly.
This method is distinguished, compared to other scheduling algorithms such as shortest-job-first and priority scheduling, that it does not suffer from starvation
and it has a balanced load on resources.
Based on analytical performance, the proposed hybrid scheduling has the markable result.
\end{abstract}

\section{Introduction}
\label{sec:intro}
Cloud Computing (CC) provides services through a set of servers that are hosted on the Internet to keep and manipulate data, rather than a on a local server. CC relies heavily on scheduling.
Optimal cloud resource allocation is needed to ensure users' satisfaction \cite{Abel}.

Scheduling is the way in which work is assigned to resources that achieve work.
Scheduling also has a role in accessing resources without excessive waiting time.
There are many types of schedulers \cite{Rahul}, such as First-Come-First-Served (FCFS), Shortest-Job-First (SJF), priority, and many more as surveyed in Section \ref{sec:Previous}.

Blockchain technology lies at the epicenter of an anticipated technological and social shock wave -- and it has already started \cite{TAPSCOTT}. Major technology companies are racing to provide solutions in this arena by providing Blockchain-as-a-service (BaaS) over the cloud. As demand for this service increases, it is expected that there will be an urgent need to schedule requests.

To enhance and provide the service uninterruptedly, this paper seeks an integration among blockchain, cloud, and a hybrid scheduling algorithm.

This paper overviews common scheduling algorithms for the cloud (Section \ref{sec:Previous}). Then it proposes the use of a hybrid scheduling algorithm with BaaS (Section \ref{sec:Proposed}). It also provides an extensive analysis of the performance of the proposed method (Section \ref{sec:Evaluation}).

\section{Related Works}
\label{sec:Previous}

Agent-based scheduling is proposed to achieve scalability and priority in real-time operation \cite{ZHU}.
Vacation queuing theory is adopted in scheduling tasks while saving energy effectively\cite{CHENG}.
%
Multi-databank scheduling is proposed based on load Divisibility \cite{SURESH}.
%
Scheduling based on meta-and hyper-heuristic is the best for reducing the tasks make span\cite{TSAI}.
Evolutionary optimization is applied to minimize tasks make span and cost \cite{ZHUb}.
%
Elastic scheduling with fault-tolerance is proposed \cite{WANG}.
%
%
A gaming cloud is proposed in \cite{ZHANG},
that balances gaming-responsiveness and costs.
While another paper proposed temporal load balancing \cite{LUO}. It succeeded to achieve cost reduction of energy too.
%

Finally, Blind Online Scheduling Algorithm (BOSA) reduces delays and
energy consumption \cite{ZHOU}. BOSA parameters are schedule operating time, user-waiting time, and performance improvement time.

Blockchain has become of great vogue in many fields \cite{DOORchain}. For instance,  it can be used as a safeguard for both security and privacy \cite{SPAINChain}.

\section{Proposed Methodology}
\label{sec:Proposed}
\begin{figure}[h]
  \centering
  \includegraphics[width=0.45\textwidth]{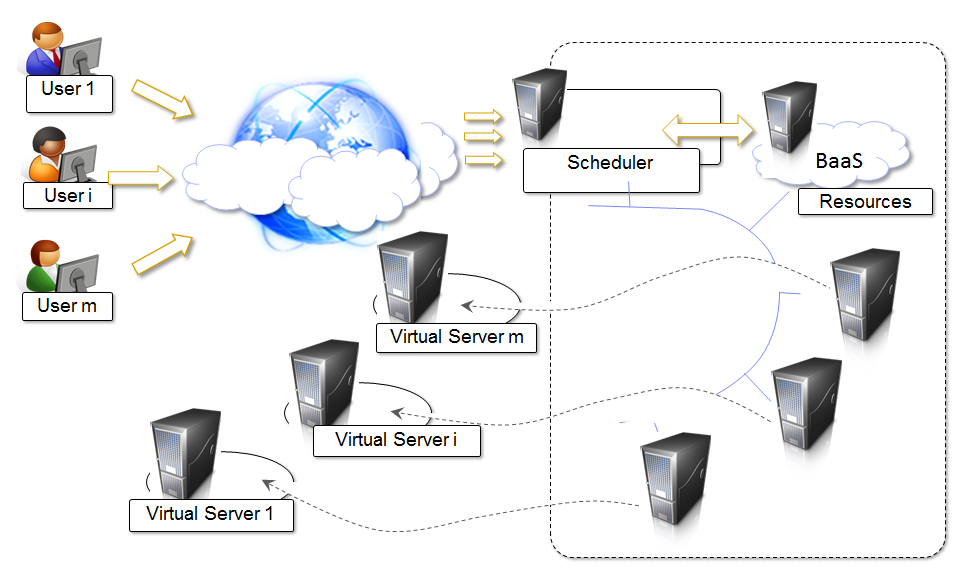}
  \caption{Proposed Methodology}\label{Proposed}
\end{figure}

This paper supposes that the data is delivered in blockchain format. This accomplishes transparency.
Suppose that there are m users
.
Each of them ($U_i$, where $1 \leq	 i \leq	 m$) asks for service from the cloud.
The scheduler receives requests,
It coordinates with the BaaS to assign a virtual machine (VM) and blockchain to each user.
The scheduler does not work once and stops, rather it continues to work as long as there are requests.
Fig~\ref{Proposed} sketches the proposed methodology.

\begin{figure}[h]
  \centering
  \includegraphics[width=0.45\textwidth]{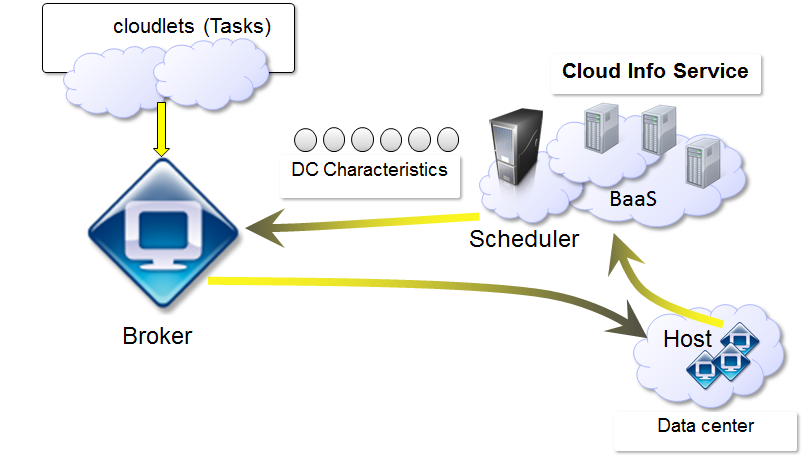}
  \caption{Components of Proposed Methodology}\label{Components}
\end{figure}

Fig~\ref{Components} sketches proposed methodology components. Proposed methodology operates as the following steps:
\begin{enumerate}
  \item Tasks (cloudlets) access the broker.
  \item The broker requests the data center(DC)
  \item DC is hosting virtual machines.
  \item DC registers in Cloud Info Service(CIS).
  \item CIS prepares the virtual blockchains.
  \item CIS sends back the DC characteristics to the broker.
\end{enumerate}
Components of proposed methodology is similar to those of
cloudSim toolkit \cite{BUYYA} which is used to model and simulate cloud environments \cite{CALHEIROS}.
Proposed methodology, however, has two extra components, namely, the BaaS layer (hosted in DC) and the scheduler (juxtaposed in CIS) .
\section{Results and Discussion}
\label{sec:Evaluation}
To evaluate the performance of the proposed methodology, the eclipse program was downloaded (Version:Neon.3 Release:4.6.3).
The cloudsim emulator was downloaded too and the settings were adjusted.

Cloudlet parameters are: length = 40000, fileSize = 300, outputSize = 300, and pesNumber = 1.

Parameters of Virtaul Machines (VMs) are: size = 10000, RAM = 512, MIPS(Millions Instructions Per Second) = 250, bandwidth = 1000, and number of CPUs (pesNumber) = 1

Setting the total number of users is done to be 1000000.

Simulation steps are as follows:
\begin{enumerate}
  \item Initialization of CloudSim package
  \item Generating Datacenters
  \item Generating Broker
  \item Generating VMs
  \item Generating Starting the simulation
  \item Printing results
\end{enumerate}

\begin{figure}[h]
  \centering
  \includegraphics[width=0.45\textwidth]{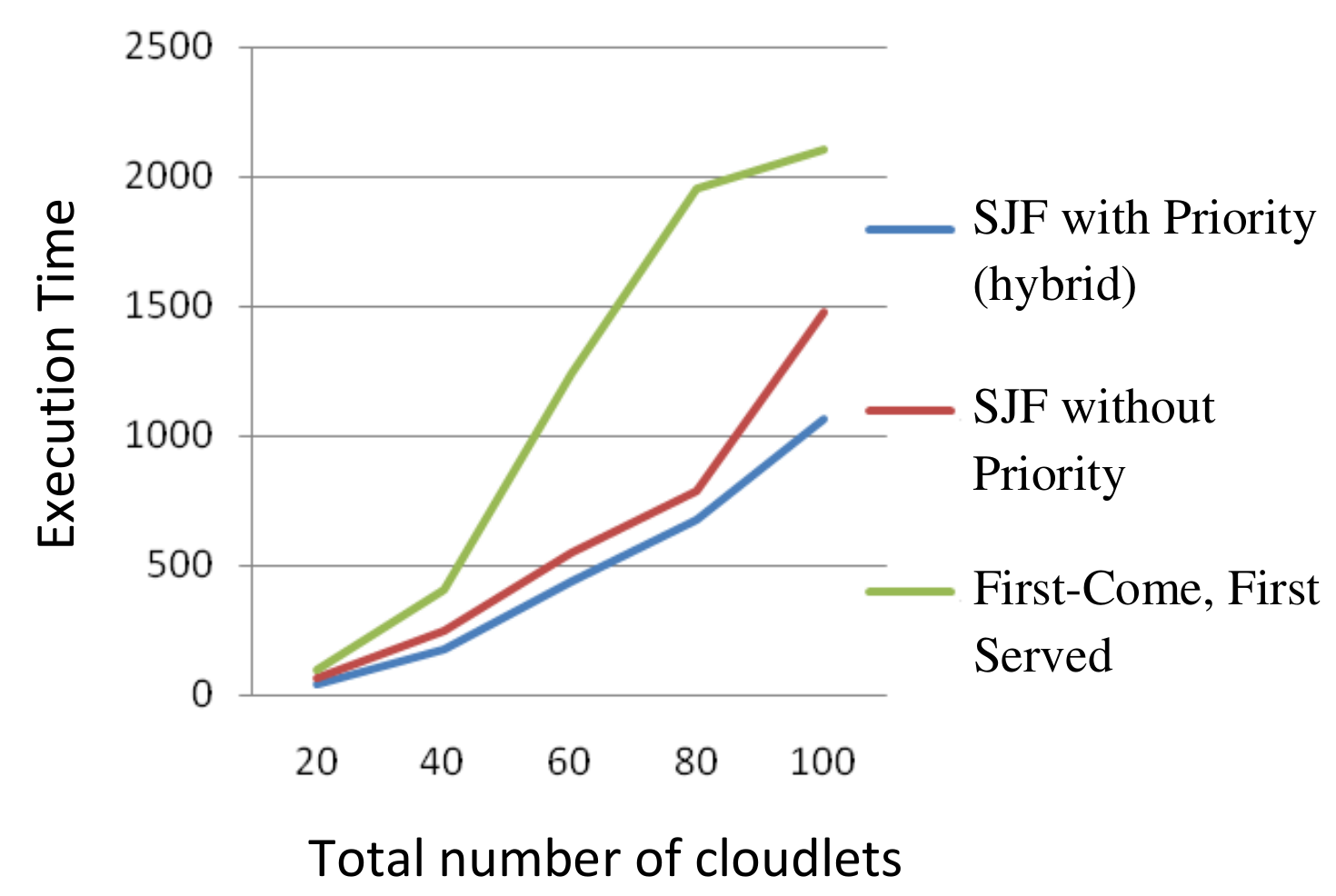}
\caption{Comparison among scheduling algorithms }
\label{Comparison}
\end{figure}

Results
are shown in Fig~\ref{Comparison}

First-Come-First-Served is the worst. The best scheduler is the hybrid of Shortest-Job-First and priority. The proposed solution does not suffer from collision or starvation.
\section{Conclusion and Future Directions}
\label{sec:Conclusion}
Scheduling is an urgent problem. It is also linked to cloud computing. Every service seeker needs it. But the criterion is to provide the service in less time.

This paper proposes the use of a hybrid scheduler with BaaS. It also provides an analysis of the performance of the proposed method.

Future directions may consider deploying the proposed methodology on any cloud service providers.


\begin{thebibliography}{6}

\bibitem{Abel}
Abel Sb. Quality Of Service-Based Resource Allocation For Web Content Delivery On Cloud Computing Infrastructure. Journal of Theoretical and Applied Information Technology. 2019 Nov 15;97(21).

\bibitem{Rahul}
Rahul M, Bansal V. A Brief Review of Scheduling Algorithms in Cloud Computing. Asian Journal of Technology and Management Research [ISSN: 2249–0892]. 2015 Jun;5(02).

\bibitem{TAPSCOTT}
TAPSCOTT, Don; TAPSCOTT, Alex. Blockchain revolution: how the technology behind bitcoin is changing money, business, and the world. Penguin, 2016.


\bibitem{ZHU}
ZHU, Xiaomin, et al. ANGEL: Agent-based scheduling for real-time tasks in virtualized clouds. IEEE Transactions on Computers, 2015, 64.12: 3389-3403.

\bibitem{CHENG}
CHENG, Chunling; LI, Jun; WANG, Ying. An energy-saving task scheduling strategy based on vacation queuing theory in cloud computing. Tsinghua Science and Technology, 2015, 20.1: 28-39.

\bibitem{SURESH}
SURESH, Sundaram; HUANG, Hao; KIM, Hyong Joong. Scheduling in compute cloud with multiple data banks using divisible load paradigm. IEEE Transactions on Aerospace and Electronic Systems, 2015, 51.2: 1288-1297.

\bibitem{TSAI}
TSAI, Chun-Wei, et al. A hyper-heuristic scheduling algorithm for cloud. IEEE Transactions on Cloud Computing, 2014, 2.2: 236-250.

\bibitem{ZHUb}
ZHU, Zhaomeng, et al. Evolutionary multi-objective workflow scheduling in cloud. IEEE Transactions on parallel and distributed Systems, 2015, 27.5: 1344-1357.

\bibitem{WANG}
WANG, Ji, et al. FESTAL: fault-tolerant elastic scheduling algorithm for real-time tasks in virtualized clouds. IEEE Transactions on Computers, 2014, 64.9: 2545-2558.

\bibitem{ZHANG}
ZHANG, Youhui, et al. A cloud gaming system based on user-level virtualization and its resource scheduling. IEEE Transactions on Parallel and Distributed Systems, 2015, 27.5: 1239-1252.

\bibitem{LUO}
LUO, Jianying; RAO, Lei; LIU, Xue. Temporal load balancing with service delay guarantees for data center energy cost optimization. IEEE Transactions on Parallel and Distributed Systems, 2013, 25.3: 775-784.

\bibitem{ZHOU}
ZHOU, Liang, et al. Exploring blind online scheduling for mobile cloud multimedia services. IEEE Wireless Communications, 2013, 20.3: 54-61.

\bibitem{DOORchain}
El-Dosuky MA, Eladl GH. DOORchain: Deep Ontology-Based Operation Research to Detect Malicious Smart Contracts. InWorld Conference on Information Systems and Technologies 2019 Apr 16 (pp. 538-545). Springer, Cham.

\bibitem{SPAINChain}
El-Dosuky MA, Eladl GH. SPAINChain: Security, Privacy, and Ambient Intelligence in Negotiation Between IOT and Blockchain. InWorld Conference on Information Systems and Technologies 2019 Apr 16 (pp. 415-425). Springer, Cham.

\bibitem{BUYYA}
BUYYA, Rajkumar; RANJAN, Rajiv; CALHEIROS, Rodrigo N. Modeling and simulation of scalable Cloud computing environments and the CloudSim toolkit: Challenges and opportunities. In: 2009 international conference on high performance computing and simulation. IEEE, 2009. p. 1-11.

\bibitem{CALHEIROS}
CALHEIROS, Rodrigo N., et al. CloudSim: a toolkit for modeling and simulation of cloud computing environments and evaluation of resource provisioning algorithms. Software: Practice and experience, 2011, 41.1: 23-50.

\bibitem{Saxena}
Deepika Saxena and R.K. Chauhan. Shortest-Job First With Fair Priority and Energy Awareness Scheduling In Green Cloud Computing.
International Journal of Trend in Research and Development, Volume 3(6), ISSN: 2394-9333, 2016.
\end{thebibliography}
\end{document}